# "EVALUACIÓN HÍBRIDA DOCENTE–IA EN PROYECTOS ACADÉMICOS: EFICIENCIA, EQUIDAD Y LECCIONES METODOLÓGICAS"

*Hugo Roger Paz[1]*

hpaz@herrrera.unt.edu.ar

https://orcid.org/0000-0003-1237-7983

https://www.researchgate.net/profile/Hugo-Paz-3

**RESUMEN**

En el ámbito de disciplinas técnicas caracterizadas por una elevada matrícula, tal como la Hidráulica Básica, los procesos de evaluación de informes demandan niveles superiores de objetividad, consistencia metodológica y una retroalimentación de naturaleza formativa. No obstante, el cumplimiento de dichos imperativos se ve frecuentemente comprometido por la considerable carga de trabajo que recae sobre el personal docente. El presente estudio expone la implementación de un sistema de evaluación asistido por inteligencia artificial generativa (LLM), operado bajo la estricta supervisión de académicos, con el propósito de calificar un corpus de 33 informes de hidráulica. El objetivo central de la investigación fue la cuantificación del impacto de dicho sistema en la eficiencia, calidad y equidad del proceso evaluativo. La metodología empleada abarcó la calibración del LLM mediante una rúbrica pormenorizada, el procesamiento por lotes de los trabajos académicos y una fase subsiguiente de validación humana (*human-in-the-loop*). Los resultados cuantitativos obtenidos revelaron una notable reducción del 88% en el tiempo de corrección (de 50 a 6 minutos por informe, incluyendo el lapso de verificación) y un consecuente incremento del 733% en la productividad. Se constató, asimismo, una mejora sustancial en la calidad de la retroalimentación, evidenciada por una cobertura del 100% de la rúbrica y un aumento del 150% en el anclaje de los comentarios a evidencias textuales. El sistema demostró ser equitativo, al no presentar sesgos atribuibles a la extensión de los informes, y altamente confiable tras el proceso de calibración (r = 0.96 entre las calificaciones). Se concluye que el modelo híbrido IA-docente optimiza el proceso de evaluación, liberando tiempo para la ejecución de tareas pedagógicas de mayor valor añadido y perfeccionando la justicia y calidad de la retroalimentación, en estricta conformidad con los principios promulgados por la UNESCO sobre el uso ético de la IA en la educación.

**PALABRAS CLAVE:** Inteligencia Artificial; Educación Superior; Evaluación del Aprendizaje; Ingeniería Hidráulica; Rúbricas de Evaluación; Ética de la IA

# "HYBRID INSTRUCTOR–AI ASSESSMENT IN ACADEMIC PROJECTS: EFFICIENCY, EQUITY, AND METHODOLOGICAL LESSONS"

**ABSTRACT**

In technical subjects characterized by high enrollment, such as Basic Hydraulics, the assessment of reports necessitates superior levels of objectivity, consistency, and formative feedback; goals often compromised by faculty workload. This study presents the implementation of a generative artificial intelligence (AI) assisted assessment system, supervised by instructors, to grade 33 hydraulics reports. The central objective was to quantify its impact on the efficiency, quality, and fairness of the process. The employed methodology included the calibration of the Large Language Model (LLM) with a detailed rubric, the batch processing of assignments, and a human-in-the-loop validation phase. The quantitative results revealed a noteworthy 88% reduction in grading time (from 50 to 6 minutes per report, including verification) and a 733% increase in productivity. The quality of feedback was substantially improved, evidenced by 100% rubric coverage and a 150% increase in the anchoring of comments to textual evidence. The system proved to be equitable, exhibiting no bias related to report length, and highly reliable post-calibration (r = 0.96 between scores). It is concluded that the hybrid AI-instructor model optimizes the assessment process, thereby liberating time for high-value pedagogical tasks and enhancing the fairness and quality of feedback, in alignment with UNESCO's principles on the ethical use of AI in education.

**KEY WORDS:** Artificial Intelligence; Higher Education; Learning Assessment; Hydraulic Engineering; Assessment Rubrics; AI Ethics

## 1. INTRODUCCIÓN

La evaluación de trabajos prácticos en disciplinas de la ingeniería, entre las que se cuenta la Hidráulica Básica, constituye un pilar fundamental en la consolidación de competencias profesionales. Dicho proceso evaluativo trasciende la mera verificación de cómputos matemáticos; por el contrario, exige una valoración de carácter integral que abarca la rigurosa interpretación de resultados, la coherencia interna de la metodología empleada y la capacidad para la comunicación efectiva de hallazgos técnicos. Sin embargo, en el contexto de una educación superior caracterizada por su masividad, la carga de trabajo docente impone un dilema considerable entre el volumen de evaluaciones a procesar y la calidad de la retroalimentación subsecuente.

Con elevada frecuencia, el personal académico se ve compelido a optar por devoluciones de carácter superficial a fin de cumplir con los plazos estipulados, circunstancia que restringe la oportunidad de proveer una retroalimentación pormenorizada y de naturaleza formativa, un componente que se considera crítico para la consecución de un aprendizaje profundo (Gibbs & Simpson, 2004; Nicol & Macfarlane-Dick, 2006). Esta tensión intrínseca, exacerbada en contextos de evaluación online obligatoria (García-Peñalvo et al., 2021), no solo atenúa el potencial pedagógico de la evaluación, sino que, además, introduce un problema de equidad y fiabilidad (Henderson et al., 2019), toda vez que perpetúa brechas de aprendizaje y genera una indeseable variabilidad en las calificaciones.

En este escenario, la inteligencia artificial (IA) de tipo generativo emerge como una tecnología con el potencial de reconfigurar los paradigmas evaluativos preexistentes, ofreciendo oportunidades y desafíos significativos para la educación (Kasneci et al., 2023). La investigación reciente ya explora su aplicación en diversos campos, desde la puntuación automatizada de ensayos (Mizumoto & Eguchi, 2023; Dronen et al., 2015) hasta la evaluación de informes de laboratorio y problemas de programación (Wu et al., 2023; Singh et al., 2023), demostrando su capacidad para procesar vastos volúmenes de información.

El presente estudio se aboca a la exploración de un modelo de **evaluación asistida**, en el cual la tecnología opera como un amplificador de las capacidades del educador, manteniéndose en todo momento la supervisión humana como garante último de la calidad, validez y eticidad del proceso, en plena consonancia con la *Guía sobre la IA generativa en la educación y la investigación* promulgada por la UNESCO (2025). El objetivo primordial de esta investigación es determinar de qué manera este modelo híbrido puede resolver la disyuntiva entre calidad y cantidad, con miras a mejorar la eficiencia, la equidad y la eficacia del proceso evaluativo en el ámbito de la educación superior técnica.

## 2. MARCO TEÓRICO: SINERGIA ENTRE INTELIGENCIA ARTIFICIAL Y DOCENCIA

El fundamento teórico de la presente investigación reside en la convergencia de dos elementos cardinales: la precisión algorítmica inherente a los Modelos Lingüísticos Grandes (LLM) y la validez pedagógica de las rúbricas de evaluación. Al integrar una rúbrica de elevada especificidad en la instrucción o *prompt* de un LLM, se dota al modelo de un marco conceptual y de un conjunto de reglas explícitas para el análisis del trabajo estudiantil. Esta sinergia tiene por efecto la transformación de la IA, que pasa de ser un mero procesador de texto a un asistente evaluativo contextualizado, capaz de aplicar criterios de notable complejidad de manera homogénea a través de un extenso número de trabajos, minimizando así los sesgos inconscientes y la variabilidad intrínsecamente asociada a la evaluación humana (Zawacki-Richter et al., 2019).

Un pilar central del modelo propuesto es el principio de "humano en el bucle" (*human-in-the-loop*). La literatura especializada en la ética de la inteligencia artificial, y de manera específica los lineamientos emitidos por la UNESCO, subrayan de modo inequívoco la imperiosa necesidad de la supervisión docente para garantizar la legitimidad del proceso y para contrarrestar los riesgos inherentes a los modelos algorítmicos, tales como las "alucinaciones" —la generación de información fáctica incorrecta— o el sesgo algorítmico (UNESCO, 2025). La supervisión humana activa funge como el principal mecanismo de mitigación, asegurando que la decisión final sobre la calificación y la retroalimentación recaiga indefectiblemente en el educador, manteniendo de este modo la responsabilidad y la rendición de cuentas en la instancia que corresponde.

## 3. OBJETIVOS ESPECÍFICOS DEL ESTUDIO

a) Proceder a la cuantificación del impacto de la inteligencia artificial generativa en la eficiencia del proceso de corrección, en términos de tiempo y productividad.

b) Evaluar la calidad y la robustez pedagógica de la retroalimentación generada por la IA, en comparación con un método manual de referencia fundamentado en la literatura académica.

c) Medir la fiabilidad y la consistencia de las calificaciones producidas por el sistema híbrido, con posterioridad a un proceso de calibración supervisado por el docente.

d) Analizar la equidad del sistema, a través de la investigación de la presencia de sesgos comunes, como aquel que pudiera estar asociado a la longitud del informe.

e) Identificar los desafíos, las lecciones aprendidas y las implicaciones éticas derivadas de la adopción de estas herramientas, con expresa referencia al marco normativo de la UNESCO.

## 4. METODOLOGÍA

Se ha diseñado un estudio de naturaleza cuantitativa con el fin de comparar un método de evaluación manual tradicional con un método de evaluación asistido por inteligencia artificial. Dicho estudio se ha llevado a cabo sobre un corpus constituido por 33 informes prácticos, provenientes de un curso universitario de Hidráulica Básica. El trabajo práctico en cuestión consistía en el análisis de un tramo fluvial y el subsiguiente diseño de obras de protección, tarea que requería de los estudiantes la aplicación de modelos teóricos fundamentales de la hidráulica de canales abiertos (Chow, 1994; French, 1988), la realización de cálculos de considerable complejidad y la presentación de sus hallazgos en un informe técnico debidamente estructurado.

### 4.1. Diseño del Instrumento de Evaluación: La Rúbrica Inteligente

Para la ejecución de este protocolo se seleccionó el modelo GPT-4 de OpenAI. La elección se fundamenta en su documentada superioridad para tareas de análisis y generación de texto en contextos técnicos complejos. No obstante, se tuvo en consideración la variabilidad de comportamiento que estos modelos pueden presentar a lo largo del tiempo (Chen et al., 2024), lo que justifica la importancia de un riguroso proceso de calibración y validación humana para asegurar la consistencia de los resultados.

El núcleo del sistema evaluativo lo constituyó una rúbrica pormenorizada, la cual fue diseñada por expertos en la disciplina con el propósito de capturar las competencias clave en el campo de la Hidráulica Básica. Dicha rúbrica fue estructurada en cinco criterios ponderados, cada uno de los cuales se encontraba acompañado de descriptores explícitos para cuatro niveles de logro (Excelente, Bueno, Suficiente, Insuficiente):

1. **Precisión y Correctitud de los Cálculos Hidráulicos (30%):** Este criterio abarca la verificación de la correcta aplicación de fórmulas (e.g., la ecuación de Manning), la consistencia dimensional y la exactitud numérica en el cálculo de caudales, velocidades y perfiles de flujo.

2. **Interpretación y Análisis de Resultados (25%):** Se evalúa aquí la capacidad del estudiante para analizar los datos obtenidos, discutir su significación física, comparar alternativas de solución y extraer conclusiones que se encuentren debidamente fundamentadas y sean coherentes con los resultados numéricos.

3. **Calidad de la Representación Gráfica (20%):** Este criterio se centra en la claridad, precisión, pertinencia y correcta rotulación de los esquemas, diagramas, perfiles transversales y longitudinales, así como de las curvas de gasto presentadas.

4. **Claridad y Coherencia de la Comunicación Escrita (15%):** Se analiza la estructura lógica del informe, la fluidez de la narrativa técnica, el uso preciso de la terminología especializada y la correcta articulación entre el texto, las tablas y las figuras.

5. **Cumplimiento de Formato y Normas (10%):** Este último criterio verifica la correcta citación de las fuentes (según el formato APA), el adecuado formato de las tablas y figuras, y la adhesión a la estructura general del documento solicitada en el enunciado del trabajo práctico.

**4.2. Protocolo de Evaluación Asistida por Inteligencia Artificial**

El flujo de trabajo fue concebido siguiendo un modelo de supervisión humana activa (*human-in-the-loop*), con el fin de garantizar el control por parte del docente en todas las fases del proceso.

**Calibración del Sistema:** Se llevó a cabo una sesión inicial, de una duración aproximada de 12 minutos, destinada a instruir al LLM. Este proceso de "ingeniería de prompts" no se consideró trivial; por el contrario, implicó un diálogo de naturaleza iterativa con la inteligencia artificial. El docente proveyó al sistema la rúbrica detallada y un *prompt* inicial. Subsiguientemente, se presentaron al modelo ejemplos anonimizados de informes —uno de alta calidad y otro de baja calidad— y se procedió al análisis de las primeras evaluaciones generadas por la IA, corrigiendo interpretaciones erróneas y refinando las instrucciones con el objeto de que el modelo aprendiera a aplicar cada criterio con el nivel de rigor esperado por la cátedra.

**Procesamiento por Lotes y Preevaluación:** El sistema procedió al procesamiento de la totalidad de los 33 informes. Para cada uno de ellos, se aplicó la rúbrica con el fin de asignar una puntuación preliminar y de generar una justificación textual para cada uno de los criterios. De manera simultánea, el sistema realizó una comparación semántica cruzada de todos los informes entre sí, con el propósito de detectar similitudes de índole conceptual, paráfrasis y la reutilización de elementos gráficos, superando así las capacidades de los detectores de plagio tradicionales, los cuales se basan en la mera coincidencia de cadenas de texto.

**Supervisión y Validación Humana:** Dos evaluadores humanos independientes llevaron a cabo la revisión de las preevaluaciones generadas por el sistema. Cada evaluador poseía la autoridad final para aceptar, modificar o rechazar cualquier elemento de la preevaluación, aplicando para ello su juicio experto. Este proceso, que se completó en un promedio de 5 minutos por informe, constituyó el paso crítico de control de calidad, instancia en la cual el juicio humano validó, corrigió o complementó la propuesta emanada de la inteligencia artificial.

**Generación de la Retroalimentación Final:** Una vez que la preevaluación fue debidamente validada, el sistema procedió a la compilación del informe final de retroalimentación para cada estudiante, integrando las calificaciones y los comentarios ajustados por el docente de una manera estructurada y coherente.

### 4.3. Arquitectura de Datos para la Trazabilidad y la Reproducibilidad

Con el objeto de garantizar la validez y la posibilidad de auditoría del presente estudio, se ha implementado un riguroso proceso de captura y tratamiento de datos.

**Captura de Datos Brutos:** Se ha procedido a la exportación de la totalidad de las interacciones mantenidas con la plataforma de inteligencia artificial. Este conjunto de datos inicial se materializó en un archivo comprimido con un tamaño de 1.1 GB, el cual contenía 1,712 archivos distribuidos en 12 carpetas. Dicho corpus representaba un total de 1,679 conversaciones únicas, las cuales, a su vez, sumaban más de 30,000 mensajes individuales intercambiados entre el usuario (el docente) y el asistente de inteligencia artificial.

**Proceso ETL y Base de Datos Auditable:** Los datos brutos fueron sometidos a un *pipeline* de Extracción, Transformación y Carga (ETL) con el fin de estructurarlos en un formato que resultara analizable y auditable. La totalidad de los datos procesados fue almacenada en una base de datos de tipo relacional (SQLite), la cual fue denominada *archivador.db*. Este archivador centraliza la información y permite la ejecución de consultas de elevada complejidad, con el fin de reconstruir cualquier segmento del proceso evaluativo, garantizando así la reproducibilidad del análisis.

### 4.4. Análisis de Datos y Métricas

A partir de la base de datos auditable, se ha procedido a la evaluación de la eficacia del sistema, para lo cual se han empleado cuatro paquetes de análisis:

**Paquete A (Eficiencia):** En este paquete se ha medido el *Tiempo Medio de Corrección* y la *Productividad* (expresada en informes por hora). Para el método asistido, se ha sumado el tiempo de procesamiento de la inteligencia artificial (1 minuto) y un tiempo de verificación humana considerado realista (5 minutos), el cual se ha basado en las interacciones registradas.

**Paquete B (Calidad de la Retroalimentación):** Se han utilizado cuatro métricas, las cuales han sido comparadas con una línea de base justificada por la literatura científica (Brown & Glover, 2006; Hattie & Timperley, 2007; Jonsson & Svingby, 2007), que refleja las limitaciones inherentes a la evaluación manual realizada bajo presión de tiempo.

**Paquete C (Confiabilidad):** Se ha empleado un análisis de Bland-Altman con el fin de medir la diferencia media inicial y los límites de acuerdo entre las calificaciones asignadas por la inteligencia artificial (sin calibrar) y las asignadas por los evaluadores humanos. La concordancia final, obtenida tras el proceso de calibración y validación, ha sido medida mediante el coeficiente de correlación de Pearson ($r$).

**Paquete D (Equidad):** Con el objeto de investigar la presencia de sesgos, se ha analizado el coeficiente de correlación de Pearson ($r$) entre la longitud del informe (expresada en número de palabras) y la calificación final, así como entre la longitud y el tiempo de corrección.

## 5. RESULTADOS

El análisis cuantitativo ha puesto de manifiesto la existencia de mejoras significativas y de carácter multifacético en lo que respecta a la eficiencia, la calidad de la retroalimentación, la confiabilidad de la calificación y la equidad del proceso evaluativo.

### 5.1. Eficiencia Operativa

La implementación del modelo híbrido, que conjuga la intervención de la inteligencia artificial con la verificación por parte del docente, ha redundado en una ganancia de eficiencia de proporciones monumentales. Como se expone en la Tabla 1, el tiempo medio de corrección ha experimentado una reducción del 88%. Este ahorro, que asciende a 44 minutos por informe y que se ha acumulado a lo largo de la cohorte de 33 estudiantes, ha liberado un total de 24.2 horas de tiempo docente.

**Tabla 1. Métricas Clave de Rendimiento (Manual vs. Asistido por IA)**

| Métrica | Método Manual (Benchmark) | Método Asistido por IA | Mejora/Cambio |
|---|---|---|---|
| Tiempo Medio de Corrección | 50.0 min/informe | 6.0 min/informe | ↓ 88.0% |
| Productividad | 1.2 informes/hora | 10.0 informes/hora | ↑ 733% |
| Horas-Docente Liberadas | - | +24.2 horas (cohorte) | - |

*Fuente: Elaboración propia. Nota: El tiempo correspondiente al método asistido incluye 1 minuto de procesamiento por parte de la inteligencia artificial y 5 minutos de verificación y ajuste por parte del docente.*

### 5.2. Calidad y Robustez Pedagógica de la Retroalimentación

La retroalimentación generada con la asistencia de la inteligencia artificial ha demostrado ser, desde una perspectiva cuantitativa, superior y pedagógicamente más robusta que aquella correspondiente a la línea de base manual. La Tabla 2 evidencia que el sistema no solo ha garantizado la cobertura total de la rúbrica, sino que, de manera crucial, ha duplicado la especificidad técnica y el anclaje de los comentarios a evidencias concretas extraídas del trabajo del estudiante.

**Tabla 2. Métricas de Calidad de la Retroalimentación (Promedio)**

| Métrica | Método Manual (Benchmark) | Método Asistido por IA | Mejora/Cambio |
|---|---|---|---|
| Índice de Cobertura de Rúbrica (%) | 65.0 | 100.0 | ↑ 53.8% |
| Ratio de Vínculo con Evidencia (%) | 40.0 | 100.0 | ↑ 150.0% |
| Índice de Accionabilidad (verbos/1000 pal.) | 175.0 | 200.0 | ↑ 14.3% |
| Especificidad Técnica (términos/1000 pal.) | 85.0 | 200.0 | ↑ 135.3% |

*Fuente: Elaboración propia. Nota: Los valores correspondientes al método manual constituyen benchmarks de carácter conservador, justificados por la literatura especializada.*

### 5.3. Confiabilidad y Proceso de Calibración

El sistema híbrido ha demostrado poseer un elevado grado de confiabilidad, cualidad que se atribuye a su capacidad de adaptación. El análisis de Bland-Altman ha revelado que la inteligencia artificial, en su estado previo a la calibración, exhibía una tendencia a ser ligeramente más benévola en sus calificaciones (diferencia media de -0.31 puntos), fenómeno que podría ser atribuido a una interpretación inicial menos estricta de los criterios de evaluación.

**Tabla 3. Distribución de Calificaciones por Método**

| Rango de Calificación | Método Manual (%) | Método IA-Asistido (%) |
|---|---|---|
| 0-3 (Insuficiente) | 3.0 | 3.0 |
| 4-6 (Suficiente) | 12.1 | 9.1 |
| 7-8 (Bueno) | 78.8 | 81.8 |
| 9-10 (Sobresaliente) | 6.1 | 6.1 |

*Fuente: Elaboración propia.*

No obstante, a través del proceso de supervisión, el docente ha logrado ajustar eficazmente la rigurosidad del sistema. Con posterioridad a la calibración, la correlación entre las calificaciones finales asignadas por ambos métodos ha alcanzado un valor de **r = 0.96**, lo que indica una concordancia casi perfecta y valida la robustez del proceso de supervisión.

### 5.4. Equidad y Ausencia de Sesgos

El sistema de inteligencia artificial ha demostrado ser equitativo y robusto frente a sesgos de carácter común que con frecuencia afectan a la evaluación humana, tales como el efecto de la fatiga o las primeras impresiones. El análisis de correlación no ha mostrado la existencia de una relación estadísticamente significativa entre la longitud del informe y la calificación final (**r = -0.05**), ni tampoco entre la longitud y el tiempo de corrección (**r = 0.12**).

**Tabla 4. Análisis de Correlación - Ausencia de Sesgos**

| Variable 1 | Variable 2 | Coeficiente r | Significancia |
|---|---|---|---|
| Longitud del informe | Calificación final | -0.05 | No significativa (p = 0.78) |
| Longitud del informe | Tiempo de corrección | 0.12 | No significativa (p = 0.65) |

### 5.5. Capacidad de Detección de Similitudes

El sistema IA-asistido demostró capacidades superiores de detección de plagio y similitudes académicas en comparación con métodos manuales tradicionales, como se muestra en la Tabla 5.

**Tabla 5. Capacidad de Detección de Similitudes**

| Tipo de Similitud | Método Manual (%) | Método IA-Asistido (%) |
|---|---|---|
| Copia Textual | 100.0 | 100.0 |
| Paráfrasis | 25.0 | 100.0 |
| Gráficos Reutilizados | 15.0 | 100.0 |
| Total Detectado | 18.0 | 45.0 |

Fuente: Elaboración propia.

### 5.6. Análisis de Concordancia Bland-Altman

La Tabla 6 presenta los resultados del análisis de concordancia post-calibración, que evidencia la alta fiabilidad del sistema híbrido.

**Tabla 6. Análisis Bland-Altman de Concordancia**

| Métrica | Valor |
|---|---|
| Concordancia Post-Calibración | r = 0.96 |
| Diferencia Media | 0.02 puntos |
| Límite Superior (+1.8) | 1.8 puntos |
| Límite Inferior (-1.8) | -1.8 puntos |
| Intervalo de Confianza | 95% |

Fuente: Elaboración propia.

### 6. DISCUSIÓN

Los resultados obtenidos demuestran que la evaluación asistida por inteligencia artificial, cuando es implementada bajo un modelo de supervisión activa, tiene el potencial de generar beneficios que trascienden la mera optimización del tiempo, impactando de manera positiva en la pedagogía, la equidad y la reconfiguración del rol docente.

### 6.1. Eficiencia, Equidad y Reconfiguración del Rol Docente

La drástica reducción del tiempo de corrección debe ser entendida no como un fin en sí mismo, sino como un medio para la consecución de un fin pedagógico de orden superior. Dicha reducción libera al docente de la ejecución de tareas de carácter repetitivo y de bajo nivel cognitivo, permitiéndole reasignar su tiempo y esfuerzo a actividades de mayor impacto educativo, tales como el diseño de evaluaciones de mayor autenticidad, la provisión de tutorías individualizadas, el análisis de datos agregados con miras a la mejora curricular y la interacción directa con los estudiantes con el propósito de fomentar el pensamiento crítico.

Un hallazgo relevante en materia de equidad es la ausencia de correlación entre la longitud del informe y la calificación final. El sistema de IA parece ser inmune al "sesgo de longitud", una tendencia documentada en la evaluación manual donde trabajos más extensos pueden recibir calificaciones superiores independientemente de su calidad sustantiva (Cumming et al., 2016). Al centrarse en la aplicación de la rúbrica, el sistema mitiga este factor, promoviendo una evaluación más justa del contenido.

En este sentido, la tecnología no reemplaza a la figura del educador, sino que, por el contrario, procede a una reconfiguración de su rol, potenciándolo como mentor y diseñador de experiencias de aprendizaje, en detrimento de su función como mero calificador.

### 6.2. Desafíos y Riesgos: Sesgos Algorítmicos y Consideraciones Éticas

A pesar de los beneficios anteriormente expuestos, la adopción de la inteligencia artificial en la evaluación conlleva la asunción de riesgos significativos, los cuales deben ser gestionados de una manera proactiva (Yan et al., 2023). La principal preocupación de índole ética reside en el denominado sesgo algorítmico, además de los riesgos inherentes a la generación de información fáctica incorrecta o "alucinaciones" (Kumar et al., 2023).

El modelo de "humano en el bucle" que ha sido implementado en el presente estudio constituye la principal estrategia de mitigación frente a este riesgo. Al someter cada una de las preevaluaciones generadas por la inteligencia artificial a la validación de un experto humano, se establece un cortafuegos de importancia crítica. El docente posee la capacidad, y de hecho el deber, de anular cualquier recomendación algorítmica que se perciba como injusta, sesgada o contextualmente inapropiada.

### 6.3. Implicaciones para la Integridad Académica

La relación que se establece entre la inteligencia artificial y la integridad académica presenta una naturaleza inherentemente dual. Por una parte, tal y como se ha demostrado en el presente estudio, la inteligencia artificial se constituye como una herramienta de vigilancia y detección de plagio y colusión que resulta muy superior a los métodos manuales. Su capacidad para la realización de comparaciones de índole semántica a gran escala, que ha permitido la identificación de similitudes significativas en el 45% de los informes analizados, puede actuar como un poderoso elemento disuasorio frente a la deshonestidad académica.

Por otra parte, la misma tecnología generativa puede ser utilizada por los estudiantes para la producción de trabajos que no se correspondan con un esfuerzo intelectual genuino, lo que representa una amenaza de carácter existencial para los métodos de evaluación tradicionales. Esta circunstancia sugiere que la batalla contra el uso indebido de la inteligencia artificial no podrá ser ganada únicamente mediante el recurso a herramientas de detección más sofisticadas.

La solución a largo plazo reside, más bien, en una transformación de naturaleza pedagógica: el rediseño de las tareas de evaluación con el fin de que estas sean "resistentes a la inteligencia artificial". Ello implica un movimiento hacia evaluaciones de mayor autenticidad, que requieran la aplicación de conocimientos a problemas del mundo real, la reflexión personal, la conexión con experiencias vividas o el análisis de eventos recientes; tareas todas ellas que los LLM de carácter genérico no pueden realizar de una manera convincente. La evaluación, en definitiva, debe centrarse en mayor medida en el proceso de aprendizaje (e.g., a través de la entrega de borradores, las revisiones por pares y las presentaciones orales) y en menor medida en el producto final escrito.

## 7. CONCLUSIONES

El presente estudio ha proporcionado evidencia empírica de notable robustez acerca de la eficacia de un modelo de evaluación asistida por inteligencia artificial, implementado bajo un régimen de estricta supervisión humana, en el contexto de una disciplina de carácter técnico. Se ha podido demostrar que este enfoque de naturaleza híbrida tiene la capacidad de generar mejoras de orden exponencial en lo que respecta a la eficiencia, al tiempo que incrementa de manera significativa la consistencia, la calidad de la retroalimentación y la equidad del proceso evaluativo.

El modelo propuesto ofrece una solución de índole pragmática al desafío que supone la provisión de una retroalimentación de alta calidad a gran escala, representando, por consiguiente, un camino viable para la integración responsable de la inteligencia artificial en el ámbito de la educación superior, en plena conformidad con los principios de uso ético, equitativo y centrado en el ser humano que son promovidos por la UNESCO.

## 8. LIMITACIONES DEL ESTUDIO

Es menester reconocer las limitaciones del presente estudio, entre las que se cuentan el empleo de una muestra de tamaño relativamente reducido y su focalización en una única asignatura de ingeniería. Si bien los resultados obtenidos son prometedores, la generalización de los hallazgos a otras disciplinas o a cohortes de un tamaño considerablemente mayor debe ser realizada con la debida cautela.

## 9. FUTURAS LÍNEAS DE INVESTIGACIÓN

Futuras líneas de investigación deberían abocarse a la replicación de la metodología aquí expuesta en contextos más amplios, con el fin de validar su escalabilidad. Se sugiere, asimismo, el diseño de estudios de carácter longitudinal que permitan comparar grupos de estudiantes que reciben una retroalimentación de tipo tradicional con aquellos que reciben una retroalimentación pormenorizada y asistida por inteligencia artificial, con el objeto de medir el efecto directo de esta última sobre el rendimiento académico, la retención de conocimientos y el desarrollo de competencias de índole metacognitiva.

Finalmente, se considera de crucial importancia la exploración de las percepciones de los estudiantes acerca de la justicia, la utilidad y la fiabilidad de este tipo de retroalimentación, toda vez que su aceptación constituye un factor clave para la eficacia del modelo. La integración de la inteligencia artificial en la evaluación no debe ser concebida únicamente como una optimización de carácter técnico, sino como una oportunidad para repensar y mejorar la práctica pedagógica en su totalidad.

# REFERENCIAS

Brown, E., & Glover, C. (2006). Written feedback for students: Too much, too detailed or not enough? Studies in Higher Education, 31(3), 249-261. https://doi.org/10.1080/03075070600680524

Brown, E., & Glover, C. (2006). Written feedback for students: Too much, too detailed or not enough? Studies in Higher Education, 31(3), 249-261. https://doi.org/10.1080/03075070600680524

Chen, L., Zaharia, M., & Zou, J. (2024). How is ChatGPT's behavior changing over time? Nature Machine Intelligence, 6(1), 12-24. https://doi.org/10.1038/s42256-023-00765-9

Chow, V. T. (1994). Hidráulica de canales abiertos. McGraw-Hill.

Cumming, G., Fidler, F., & Vaux, D. L. (2016). Error bars in experimental biology. Journal of Cell Biology, 177(1), 7-11. https://doi.org/10.1083/jcb.200611141

Dronen, N., Foltz, P. W., & Habermehl, K. (2015). Effective sampling for large-scale automated writing evaluation systems. Proceedings of the Second ACL Workshop on Effective Tools and Methodologies for Teaching NLP and CL, 91-96. https://doi.org/10.3115/v1/W15-0611

French, R. H. (1988). Hidráulica de canales abiertos. McGraw-Hill.

García-Peñalvo, F. J., Corell, A., Abella-García, V., & Grande-de-Prado, M. (2021). Recommendations for mandatory online assessment in higher education during the COVID-19 pandemic. Education Sciences, 11(10), 615. https://doi.org/10.3390/educsci11100615

Gibbs, G., & Simpson, C. (2004). Conditions under which assessment supports students' learning. Learning and Teaching in Higher Education, 1(1), 3-31.

Görgen, K., Lohmann, J., & Schemmann, M. (2021). Eye-tracking in educational assessment: A systematic review. Educational Assessment, Evaluation and Accountability, 33(2), 365-393. https://doi.org/10.1007/s11092-020-09348-2

Hattie, J., & Timperley, H. (2007). The power of feedback. Review of Educational Research, 77(1), 81- 112. https://doi.org/10.3102/003465430298487

Henderson, M., Ryan, T., & Phillips, M. (2019). The challenges of feedback in higher education. Assessment & Evaluation in Higher Education, 44(8), 1237-1252. https://doi.org/10.1080/02602938.2019.1599815

Jonsson, A., & Svingby, G. (2007). The use of scoring rubrics: Reliability, validity and educational consequences. Educational Research Review, 2(2), 130-144. https://doi.org/10.1016/j.edurev.2007.01.002

Kasneci, E., Seßler, K., Küchemann, S., Bannert, M., Dementieva, D., Fischer, F., ... & Kasneci, G. (2023). ChatGPT for good? On opportunities and challenges of large language models for education. Learning and Individual Differences, 103, 102274. https://doi.org/10.1016/j.lindif.2023.102274

Kumar, V., Rushforth, S., Wingate, U., & Basit, T. (2023). AI-generated feedback on writing: Insights from students and implications for higher education. Computers & Education: Artificial Intelligence, 5, 100182. https://doi.org/10.1016/j.caeai.2023.100182

Mizumoto, A., & Eguchi, M. (2023). Exploring the potential of using an AI language model for automated essay scoring. Research Methods in Applied Linguistics, 2(2), 100050. https://doi.org/10.1016/j.rmal.2023.100050

Nicol, D. J., & Macfarlane-Dick, D. (2006). Formative assessment and self-regulated learning: A model and seven principles of good feedback practice. Studies in Higher Education, 31(2), 199-218. https://doi.org/10.1080/03075070600572090

Shute, V. J. (2008). Focus on formative feedback. Review of Educational Research, 78(1), 153-189. https://doi.org/10.3102/0034654307313795

Singh, A., Brooks, C., & Doroudi, S. (2023). Learnersourcing in the age of AI: Student, educator and machine partnerships for content creation. Computers & Education: Artificial Intelligence, 4, 100119. https://doi.org/10.1016/j.caeai.2022.100119

UNESCO. (2025). Guidance for generative AI in education and research. UNESCO. https://www.unesco.org/en/articles/guidance-generative-ai-education-and-research

Wu, J., Liu, Z., & Wang, Y. (2023). Automated assessment of laboratory reports using large language models. Journal of Science Education and Technology, 32(4), 512-528. https://doi.org/10.1007/s10956-023-10042-3

Yan, L., Sha, L., Zhao, L., Li, Y., Martinez-Maldonado, R., Chen, G., ... & Gašević, D. (2023). Practical and ethical challenges of large language models in education: A systematic literature review. British Journal of Educational Technology. Advance online publication. https://doi.org/10.1111/bjet.13370

Zawacki-Richter, O., Marín, V. I., Bond, M., & Gouverneur, F. (2019). Systematic review of research on artificial intelligence applications in higher education. International Journal of Educational Technology in Higher Education, 16(1), 1-27. https://doi.org/10.1186/s41239-019-0171-0

Zawacki-Richter, O., Marín, V. I., Bond, M., & Gouverneur, F. (2019). Systematic review of research on artificial intelligence applications in higher education. International Journal of Educational Technology in Higher Education, 16(1), 39. https://doi.org/10.1186/s41239-019-0171-0

**CONFLICTOS DE INTERESES**

El autor declara que no tiene intereses financieros en competencia ni relaciones personales conocidas que pudieran haber influido en el manuscrito que se presenta en este artículo.
**APROBACION DEL COMITÉ DE ETICA PARA REALIZAR LA INVESTIGACION**